\documentclass[aps,showpacs,manuscript,12pt]{revtex4}
\usepackage{amssymb}
\usepackage{amsmath}
\usepackage{graphicx}

\begin{document}

\title{\textbf{On the initial conditions of the 1-point PDF for incompressible
Navier-Stokes fluids}}
\author{M. Tessarotto$^{1,2}$ and C. Asci$^{1}$}
\affiliation{$^{1}$Department of Mathematics and Informatics,\\
University of Trieste, Trieste, Italy \\
$^{2}$ Consortium for Magnetofluid Dynamics, Trieste, Italy }

\begin{abstract}
An aspect of fluid dynamics lies in the search of possible
statistical models for Navier-Stokes (NS) fluids described by
classical solutions of the incompressible Navier-Stokes equations
(INSE). This refers in particular to statistical models based on
the so-called inverse kinetic theory (IKT) . This approach allows
the description of fluid systems by means a suitable 1-point
velocity probability density function (PDF) which determines, in
terms of suitable "moments", \textit{the complete set of fluid
fields} which define the fluid state. A fundamental related issue
lies in the problem of the \textit{unique construction of the
initial PDF}. The goal of this paper is to propose a solution
holding for NS fluids. Our claim is that the initial PDF  can be
uniquely determined by imposing a suitable set of {\it physical
realizability constraints}.
\end{abstract}

\date{\today }
\maketitle

\pagebreak

\section{Introduction: CSM-inspired statistical models}

Fundamental aspects of fluid dynamics are related to construction of
statistical models $\left\{ \ f,\Gamma \right\} $ for fluid systems. These
are sets $\left\{ f,\Gamma \right\} $ formed by a suitable probability
density function (PDF) and a phase-space $\Gamma $ (subset of $%
\mathbb{R}
^{n}$) on which $f$ is defined.\ By definition, a statistical model $\left\{
f,\Gamma \right\} $ of this type must permit the representation, via a
suitable mapping%
\begin{equation}
\left\{ f,\Gamma \right\} \Rightarrow \left\{ Z\right\} ,
\end{equation}%
of the \textit{fluid fields} $\left\{ Z\right\} \equiv \left\{
Z_{i},i=1,n\right\} $ which define the state of the same fluids. Depending
whether the mapping provides \textit{the complete set }or only\textit{\ a
subset} of $\left\{ Z\right\} $ the statistical model is denoted
respectively \textit{complete} or \textit{incomplete}. In the case of a
Navier-Stokes (NS) fluid, i.e., an incompressible isothermal and isentropic
Newtonian fluid described by the incompressible NS equations (INSE), the
fluid fields are
\begin{equation}
\left\{ Z\right\} \equiv \left\{ \rho _{0},\mathbf{V},p_{1},S_{T}\right\} ,
\label{fluid fields}
\end{equation}%
where in particular $\rho _{0}$ (the mass density) and $S_{T}$ (the
thermodynamic entropy) are both assumed constant in $\overline{\Omega }%
\times I$ (with $\overline{\Omega }$ denoting the closure of the
configuration domain $\Omega \subseteq
\mathbb{R}
^{3}$ and $I\subseteq
\mathbb{R}
$ a time interval/axis). Moreover, $\mathbf{V}$ and $p_{1}$ denote the fluid
velocity and the (strictly positive) kinetic pressure
\begin{equation}
p_{1}(\mathbf{r},t)=p(\mathbf{r},t)+p_{0}(t)+\phi (\mathbf{r},t),
\label{p_1}
\end{equation}%
where $p(\mathbf{r},t),p_{0}(t)$ and $\phi (\mathbf{r},t)$ represent
respectively the fluid pressure, the (strictly-positive) pseudo-pressure and
the (possible) potential associated to the conservative volume force density
acting on the fluid. The construction of a statistical model involves the
introduction of a suitable PDF of some sort ($f$), defined on an appropriate
phase-space $\Gamma $ in terms of which the fluid fields (or only a proper
subset of them) can be represented via suitable \textit{velocity} or \textit{%
phase-space moments} of the PDF. \ A well-known example of incomplete
statistical model, holding for incompressible NS fluids, is provided by the
so-called \textit{statistical hydromechanics} developed originally by Hopf
\cite{Hopf1952} and later extended by Rosen \cite{Rosen1960} and Edwards
\cite{Edwards1964} (\textit{HRE approach}), based on the statistical model $%
\left\{ f_{H},\Gamma _{1}\right\} $. This relies on the introduction of the $%
1-$\textit{point (or local) velocity-space PDF, }$f_{1},$ to be intended as
the conditional PDF of the velocity $\mathbf{v}$ (\textit{kinetic velocity})
with respect the remaining variables. In the HRE approach these are
identified with $(\mathbf{r,}t),$ where $(\mathbf{r,}t)\in $ $\Omega \times
I,$ while $f_{1}\equiv f_{1}(\mathbf{r,u,}t;Z),$ with $\mathbf{u\equiv v-V}(%
\mathbf{r,}t)$ the relative kinetic velocity$,$ is identified with
\begin{equation}
f_{H}\equiv \delta \left( \mathbf{v-V}(\mathbf{r,}t)\right)
\label{DIRAC DELTA- HRE}
\end{equation}%
$f_{H}$ denoting the three-dimensional Dirac delta defined in the velocity
space $U\subseteq
\mathbb{R}
^{3},$ with\ $\mathbf{v}$ belonging to $U$ and $(\mathbf{r,}t)\in \Omega
\times I$ (with $\Omega \subseteq
\mathbb{R}
^{3}$ and $I$ the configuration domain and the time axis). Hence, $f_{H}$ is
defined by assumption on the set spanned by the state vector $\mathbf{x=}(%
\mathbf{\mathbf{r,v}})\mathbf{,}$ i.e.,  the \textit{restricted phase space }%
$\Gamma _{1}\equiv \Omega \times U$ $.$ It follows that, in this case,
\textit{only one velocity moment} of $f_{1},$ corresponding to $G=\mathbf{v,}
$ is actually prescribed in terms of the fluid fields and reads
\begin{equation}
\int\limits_{U}d^{3}\mathbf{v}Gf_{H}(\mathbf{r,u,}t;Z)=\mathbf{V}(\mathbf{r,}%
t).  \label{moments of f_1 -HRE APPROACH}
\end{equation}

The statistical model $\left\{ f_{H},\Gamma _{1}\right\} $, which
was adopted also by to Monin \cite{Monin1967} and Lundgren
\cite{Lundgren1967}, belongs actually to a more general class of
statistical models inspired by Classical Statistical Mechanics
(CSM).  In this context, a convenient alternative approach - which
permits instead the representation in terms of $f$ of the complete
set of fluid fields (hence providing a \textit{complete
statistical model}) - is provided by \ with the so-called
\textit{IKT-statistical model }$\left\{ f_{1},\Gamma _{1}\right\}
,$ developed in the framework of inverse kinetic theory (IKT;
Tessarotto \textit{et al.} \cite%
{Tessarotto2004,Ellero2005,Tessarotto2006,Tessarotto2007b,Tessarotto2008-4}).
This is based on the construction of a suitable 1-point velocity-space PDF, $%
f_{1},$\ defined in such a way to yield,\textit{, in terms of a prescribed
set of velocity moments of the same PDF, the complete set of fluid fields}
characterizing a NS fluids.

A basic issue is therefore represented by the construction of the \textit{%
general solution} $f_{1}(t)$ for the IKT-statistical model$,$ and in
particular \textit{the initial} PDF $f_{1}(t_{o}),$ based on the available
information (i.e., observables) on the fluid system.

The goal of this paper is to point out that $f_{1}(t)$ can be uniquely
determined at the initial time $t=t_{o}$. The basic results are represented
by THMs 1 and 2 (see Sections 4 and 5). We claim that, subject to the
requirement of factorization at the initial time for all multi-point PDFs
the initial PDF $f_{1}(t_{o})$ can be determined imposing the principle of
entropy maximization (PEM \cite{Jaynes1957}) on the Boltzmann-Shannon
entropy \cite{Tessarotto2007}. In detail, we intend to show that the initial
PDF $f_{1}(t_{o})$ can be determined in such a way to satisfy the following
\textit{requirements}:\

\begin{enumerate}
\item \textit{(requirement \#1) }it satisfies \textit{the complete set of
constraints placed at the initial time }$t=t_{o}$\textit{\ by the
observables defined for the fluid system (physical realizability conditions)}%
;

\item \textit{(requirement \#2) }it does not require the specification of
higher-order PDFs, with\ $s>2;$

\item \textit{(requirement \#3) }it \textit{provides the general solution
for the initial PDF required by the} \textit{IKT-statistical model }$\left\{
f_{1},\Gamma _{1}\right\} .$
\end{enumerate}

\section{Physical observables of a NS fluids}

An important preliminary task to accomplish is to determine the complete set
of physical observables required to define the functional class of the PDF $%
\left\{ \left\langle f_{1}(t)\right\rangle _{\Omega }\right\} ,$ \textit{to
be prescribed at the initial time} $t_{o}.$ These include, in particular,
the complete set of the fluid field which specify the state of the fluid. In
the IKT approach \cite%
{Tessarotto2004,Ellero2005,Tessarotto2006,Tessarotto2007b,Tessarotto2008-4}
for an isothermal, isentropic and incompressible NS fluid the latter must be
represented in terms of the set (\ref{fluid fields}).

A further observable can, however, be defined in terms of the velocity field
$\mathbf{V}\left( \mathbf{r,}t\right) .$\ This is represented by the\textit{%
\ continuous velocity-frequency density} (C-VFD), $\widehat{f}%
_{1}^{(freq)}(t,\mathbf{v}),$ namely the velocity-frequency associated to
the fluid velocity occurring in subsets of the fluid domain $\Omega
\subseteq
\mathbb{R}
^{3},$ to be defined in particular so that%
\begin{equation}
\int\limits_{U}d^{3}v\widehat{f}_{1}^{(freq)}(t,\mathbf{v})=1,
\label{normalization}
\end{equation}%
$U\equiv
\mathbb{R}
^{3}$ denoting the three-dimensional velocity space. For definiteness, let
us assume in the remainder that:

\begin{itemize}
\item \emph{Assumption \#1:} the fluid domain $\Omega $ is a bounded subset
of $%
\mathbb{R}
^{3}$ with finite measure (for example is a cube).

\item \emph{Assumption \#2}: the fluid velocity $\mathbf{V}(\mathbf{r,}t)$
is bounded. Hence there exists a positive constant $V_{B}\in
\mathbb{R}
,$ with $V_{B}(t)=2\sup \left\{ V_{k}(\mathbf{r,}t);k=1,2,3;\mathbf{r}\in
\Omega \right\} $ and we can define the bounded subset of $%
\mathbb{R}
^{3}$
\begin{equation*}
U=\left\{ \mathbf{v:}\left\vert v_{i}\right\vert \leq \frac{1}{2}V_{B};%
\mathbf{v\in }%
\mathbb{R}
^{3}\right\} ,
\end{equation*}%
to be identified with the velocity space of the statistical model.
\end{itemize}

Let us require for simplicity that, thanks to \emph{Assumption \#1,} $\Omega
$ is partitioned in $N$ like cubic cells. Then let us introduce the notion
of \textit{continuous velocity-frequency density }(C-VFD) $\widehat{f}%
_{1}^{(freq)}(t,\mathbf{v})$ to be defined as the limit function
\begin{equation}
\widehat{f}_{1}^{(freq)}(t,\mathbf{v})=\lim_{N\rightarrow \infty }\widehat{f}%
_{1,N}^{(freq)}(t,\mathbf{v}),  \label{continuous VDF}
\end{equation}%
$\widehat{f}_{1,N}^{(freq)}(t,\mathbf{v})$ indicating the corresponding
(discrete\textit{) velocity-frequency density function }(V-FDF) defined on a
N-cell partition of $\Omega $
\begin{equation}
\widehat{f}_{1,N}^{(freq)}(t,\mathbf{v})\equiv \frac{1}{N}%
\sum\limits_{i=1,N}N_{1}(\mathbf{r}_{i}\mathbf{,v,}t).  \label{discrete VFD}
\end{equation}%
To define $N_{1}(\mathbf{r}_{i}\mathbf{,v,}t)$ let us notice that - thanks
to \emph{Assumption \#2}\ - for each component of the fluid velocity $V_{k}(%
\mathbf{r,}t)$ (with $k=1,2,3$) the inequality
\begin{equation}
\left\vert V_{k}(\mathbf{r,}t)\right\vert \leq \frac{1}{2}V_{B}  \label{B.1}
\end{equation}%
holds. Hence, the velocity space $U$ can be partitioned in $M$ like cubic
cells, with $M\in
\mathbb{N}
_{0}$ denoting an arbitrary integer. Therefore, if $\mathbf{r}_{i}$ denotes
the position of the center of mass for the $i-th$ configuration-space cell, $%
N_{1}(\mathbf{r}_{i}\mathbf{,v,}t)$ can be defined as the frequency of
occurrence of the velocity fluid field $\mathbf{v,}$ assumed to belong to
the velocity-space cell defined (for $k=1,2,3)$ by the inequalities $%
\left\vert V_{k}(\mathbf{r}_{i},t)-v_{k}\right\vert \leq \frac{V_{B}}{2M}.$
Thus, $N_{1}(\mathbf{r}_{i}\mathbf{,v,}t)$ can be defined as
\begin{eqnarray}
&&N_{1}(\mathbf{r}_{i}\mathbf{,v,}t)=\frac{1}{c}\prod\limits_{k=1,2,3}\Theta
_{ik}(\mathbf{v}),  \label{B.3} \\
&&\Theta _{ik}(\mathbf{v})\equiv \Theta (V_{k}(\mathbf{r}_{i},t)-v_{k}-\frac{%
V_{B}}{2M})  \notag \\
&&\Theta (v_{k}-V_{k}(\mathbf{r}_{i},t)+\frac{V_{B}}{2M}),  \label{B.4b}
\end{eqnarray}%
with $\Theta (x)$ denoting the Heaviside theta function, $c\in
\mathbb{R}
$ the normalization constant%
\begin{equation}
c=\int\limits_{U}d^{3}v\frac{1}{N}\sum\limits_{i=1,N}\prod\limits_{k=1,2,3}%
\Theta _{ik}(\mathbf{v})  \label{B.6}
\end{equation}%
and $M,N\in
\mathbb{N}
_{0}$ arbitrary integers. We stress that they can always be defined so that $%
M=M(N)$ is a strictly monotonic function of $N$ (in particular, $M$ and $N$
can be, for example, so that $N=M^{3}).$ Then, thanks to positions (\ref{B.3}%
)-(\ref{B.6}), by construction $\widehat{f}_{1,N}^{(freq)}(\mathbf{v,}t)$
satisfies the normalization condition
\begin{equation}
\int\limits_{U}d^{3}v\widehat{f}_{1,N}^{(freq)}(\mathbf{v,}t)=1.
\end{equation}%
This manifestly implies for the limit function $\widehat{f}_{1}^{(freq)}(%
\mathbf{v,}t)$ the analogous condition of normalization (\ref{normalization}%
).

Additional observables, to be defined in analogy to $\widehat{f}%
_{1}^{(freq)}(\mathbf{v,}t),$ are however represented by:

\begin{enumerate}
\item the $2$\textit{-point velocity difference-frequency density function}
(VD-FDF) $\widehat{f}_{2}^{freq}(\mathbf{r},\mathbf{v,}t_{o})$ to be
identified with the frequency of the velocity difference $\mathbf{v=\mathbf{V%
}_{1}-V_{2}}$ occurring between two positions $\mathbf{r}_{1}$ and $\mathbf{r%
}_{2}$ with displacement $\mathbf{r=r}_{1}-\mathbf{r}_{2}$ and subject to
the normalization%
\begin{equation}
\int\limits_{U}d^{3}v\widehat{f}_{2}^{freq}(\mathbf{r},\mathbf{v,}t_{o})=1;
\end{equation}

\item as well as the analogous $s$\textit{-point velocity-difference FDF}s
which can be defined in principle for arbitrary $s\geq 2.$
\end{enumerate}

It is obvious that in principle \textit{all the constraints provided by the
infinite set of velocity-difference PDFs should be satisfied} by the initial
1-point $f_{1}(t_{o})$ [which defines the IKT-statistical model $\left\{
f_{1},\Gamma _{1}\right\} ]!$

\section{Physical realizability conditions for $f_{1}(t_{o})$}

In this reference, a natural question arises, i.e., whether the
arbitrariness [in the definition of $\left\{ f_{1},\Gamma _{1}\right\} $]
can be used,\textit{\ by proper prescription on its functional class} $%
\left\{ \left\langle f_{1}(t_{o})\right\rangle _{\Omega }\right\} ,$\textit{%
\ to determine it uniquely consistent not only with INSE but also with the
relevant physical observables. \ }An important preliminary task to
accomplish is to establish \ the relationship of $f_{1}$ with the fluid
fields. More precisely, here we state that\textit{, besides the complete set
of fluid fields evaluated at the initial time }$\left\{ \mathbf{V}\left(
\mathbf{r,}t_{o}\right) ,p_{1}\left( \mathbf{r,}t_{o}\right)
,S_{T}(t_{o})\right\} ,$\textit{\ the PDF must also be suitably related to
the\ initial }$1-$\textit{point velocity-frequency density function }$%
\widehat{f}_{1}^{(freq)}(\mathbf{v,}t_{o}).$

In the following we shall require that $f_{1}$ satisfies the following
constraints (to be intended as \textit{physical realizability conditions}
for $f_{1}$):

\textit{\ \emph{Realizability condition \#1}: }it admits for all $\left(
\mathbf{r,}t\right) \in \overline{\Omega }\times I$ \ (including the initial
time $t_{o}$) and $G=1,\mathbf{v,}u^{2}/2,\mathbf{uu},\mathbf{u}%
u^{2}/2,lnf_{1}(\mathbf{r,v,}t)$ the velocity and phase-space moments $%
\int\limits_{U}d\mathbf{v}Gf_{1}$ and $\int\limits_{\Gamma _{1}}d\mathbf{v}%
f_{1}\ln f_{1}$ and satisfies the constraint equations (denoted as \textit{%
correspondence principle}):
\begin{eqnarray}
\int\limits_{U}d\mathbf{v}Gf_{1}(\mathbf{r,v,}t) &=&1,\mathbf{V}(\mathbf{r,}%
t),p_{1}(\mathbf{r,}t),  \label{MOMENTS} \\
S(f_{1}(t)) &=&S_{T},  \label{ENTROPY}
\end{eqnarray}%
with $S(f_{1}(t))=-\int\limits_{\Gamma _{1}}d\mathbf{x}f_{1}(\mathbf{r,v,}%
t)\ln f_{1}(\mathbf{r,v,}t)$ denoting the Boltzmann-Shannon statistical
entropy associate to $f_{1}(t)$ and $\Gamma _{1}$ the phase space $\Gamma
_{1}=\Omega \times U;$

\emph{Realizability condition \#2}\textit{:} at the initial time $t=t_{o}$
it satisfies the constraint:

\begin{equation}
\left\langle f_{1}(t_{o})\right\rangle _{\Omega }=\widehat{f}_{1}^{(freq)}(%
\mathbf{v,}t_{o}).  \label{constraint on f1 (at
initial time)}
\end{equation}%
where $f_{1}(t)\equiv f_{1}(\mathbf{r,v,}t;Z)$ and $\left\langle
f_{1}(t)\right\rangle _{\mathbf{r,}\Omega }$ denotes the $\Omega -$average
at time $t$
\begin{equation}
\left\langle a(\mathbf{r,v,}t)\right\rangle _{\mathbf{r,}\Omega }\equiv
\frac{1}{\mu (\Omega )}\int\limits_{\Omega }d^{3}\mathbf{r}a(\mathbf{r,v,}t).
\label{f_1-average}
\end{equation}

\emph{Realizability condition \#3}\textit{:} at the initial time $t=t_{o},$
we require that the 2-point PDF $f_{2}(1,2\mathbf{,}t_{o})$ satisfies the
constraint:

\begin{equation}
\widehat{f}_{2}(\mathbf{r},\mathbf{v,}t_{o})=\widehat{f}_{2}^{freq}(\mathbf{r%
},\mathbf{v,}t_{o}),
\end{equation}%
where $\widehat{f}_{2}(\mathbf{r},\mathbf{v,}t_{o})$ denotes the
velocity-difference 2-point PDF defined as
\begin{equation}
\widehat{f}_{2}(\mathbf{r},\mathbf{v,}t_{o})=\frac{1}{\mu (\Omega )}\int d%
\mathbf{R}\int d\mathbf{V}f_{2}(1,2\mathbf{,}t_{o}).
\end{equation}%
Then requiring that $f_{2}(1,2,t_{o})$ is factorized%
\begin{equation}
f_{2}(1,2,t_{o})=f_{1}(1,t_{o})f_{1}(2,t_{o})  \label{FACTORIZATION-1}
\end{equation}%
this implies the constraint%
\begin{equation}
\frac{1}{\mu (\Omega )}\int d\mathbf{R}\int d\mathbf{V}%
f_{1}(1,t_{o})f_{1}(2,t_{o})=\widehat{f}_{2}^{freq}(\mathbf{r},\mathbf{v,}%
t_{o}).  \label{NEW CONSTRAINT-1}
\end{equation}

\section{Consequences of PEM}

Let us now show how $f_{1}(1,t_{o})\equiv f_{1}(\mathbf{x}_{1}\mathbf{,}%
t_{o};Z)$ can be determined for the IKT-statistical model $\left\{
f_{1},\Gamma _{1}\right\} $ by imposing \cite{Ellero2005,Tessarotto2006}
that at $t=t_{o}$ it satisfies the \textit{constrained maximal variational
principle} (also known as principle of entropy maximization or PEM; Jaynes,
1957 \cite{Jaynes1957} :\textit{\ \ }%
\begin{equation}
\delta S(f_{1}(1,t))=0,  \label{A-2}
\end{equation}%
together with the \textit{physical realizability conditions} defined above.
For definiteness, let us assume that $\ f_{1}(1,t_{o})$ is an ordinary,
strictly positive function of the general form%
\begin{equation}
f_{1}(1,t_{o})=\widehat{f_{1}}(1,t_{o})h(1,t_{o})\left\langle
f_{1}(1,t)\right\rangle _{\Omega }\frac{h(1,t)}{\left\langle
h(1,t)\right\rangle _{\Omega }},  \label{NON-Maxwellian}
\end{equation}%
with
\begin{equation}
\widehat{f_{1}}(1,t_{o})=\frac{\left\langle f_{1}(1,t_{o})\right\rangle
_{\Omega }}{\left\langle h(1,t_{o})\right\rangle _{\Omega }}.
\end{equation}%
$\left\langle f_{1}(1,t_{o})\right\rangle _{\Omega }$ determined by the
constraint Eq.(\ref{constraint on f1 (at initial time)})\textit{\ }and%
\textit{\ }$h(t_{o})$ to be assumed as a strictly positive and regular real
function.

Let us determine the initial PDF $f_{1}(1,t_{o})$ subject to the \textit{%
whole set of constraints} placed by the aforementioned by the physical
realizability conditions. Invoking the factorization condition (\ref%
{FACTORIZATION-1}) and the position (\ref{NON-Maxwellian}) the extremal
1-point PDF $f_{1}(1,t_{o})$ can be determined in an equivalent way either
from the variational principle (\ref{A-2}) of for the 2-point entropy
principle
\begin{equation}
\delta S(f_{2}(t_{o}))=0.  \label{PEM-2}
\end{equation}%
In fact thanks to (\ref{FACTORIZATION-1}) it follows that $S(f_{2})=-2\mu
(\Omega )\int dx_{1}f_{1}(1,t_{o})\ln f_{1}(1,t_{o})\equiv 2\mu (\Omega
)S(f_{1}).$ Hence one obtains
\begin{eqnarray}
&&\left. \delta S(f_{2})=-2\mu (\Omega )\int dx_{1}\widehat{f_{1}}%
(1,t_{o})\delta h(1,t_{o})\left[ 1+\ln h(1,t_{o})+\right. \right. \\
&&\left. \left. +\lambda _{o}(1,t_{o})+\lambda
_{2}(1,t_{o})u(1,t_{o})^{2}+\int dx_{2}\lambda _{3}(\mathbf{r},\mathbf{v,}%
t_{o})f_{1}(2,t_{o})\right] =0\right. ,  \notag
\end{eqnarray}%
where $\lambda _{3}(\mathbf{r},\mathbf{v,}t_{o})$ is a suitable Lagrange
multiplier to be determined in such a way to fulfill the constraint (\ref%
{FACTORIZATION-1}). Then, considering variations of $h(1,t_{o})$ of the form
$\delta f_{1}(1,t_{o})=\widehat{f_{1}}(1,t_{o})\delta h(1,t_{o}),$ i.e.,
defined so that there results identically $\delta \widehat{f_{1}}%
(1,t_{o})\equiv 0,$ the previous variational principle requires that $%
h(1,t_{o})$ must fulfill the variational equation
\begin{eqnarray}
&&\int\limits_{\Gamma _{1}}d\mathbf{x}_{1}\mathbf{\delta }h(1,t_{o})\frac{%
\left\langle f_{1}(1,t_{o})\right\rangle _{\Omega }}{\left\langle
h(1,t_{o})\right\rangle _{\Omega }}\left[ 1+\ln h(1,t_{o})+\lambda _{o}(%
\mathbf{r}_{1},t_{o})\right.  \label{E-L equation} \\
&&\left. \left. +\lambda _{2}(\mathbf{r}_{1},t_{o})u(1,t_{o})^{2}+\int%
\limits_{\Gamma _{1}}d\mathbf{x}_{2}\lambda _{3}(\mathbf{r},\mathbf{v,}%
t_{o})f_{1}(\mathbf{r}_{2},\mathbf{v}_{2},t_{o})\right] =0,\right.  \notag
\end{eqnarray}%
with summation understood on the index $i$ (for $i=1,2,3$). \ The result,
which provides the sought solution for $f_{1}(t_{o}),$ has the flavor of:

\vskip5mm \textbf{THM.1 - General solution for} $f_{1}(t_{o})$

\textit{The general solution of the PEM variational principle [Eq.(\ref{A-2}%
)] in the functional class }$\left\{ \left\langle f_{1}(t_{o})\right\rangle
_{\Omega }\right\} $ \textit{determined by realizability conditions \#1-\#3
is of the form (\ref{NON-Maxwellian})} \textit{with} $h(1,t_{o})$ \textit{%
taking the form of a non-Gaussian distribution}%
\begin{equation}
h(1,t_{o})=\exp \left\{ -1-\lambda _{o}(\mathbf{r}_{1},t_{o})-\lambda _{2}(%
\mathbf{r}_{1},t_{o})u(1,t_{o})^{2}-\widehat{\lambda }_{3}(1,t_{o})\right\} ,
\label{solution for h}
\end{equation}%
\textit{where} $\lambda _{o}(\mathbf{r,}t_{o})$ and $\lambda _{2}(\mathbf{r,}%
t_{o})$ \textit{suitable Lagrange multipliers to be determined imposing at} $%
t=t_{o}$ \textit{the moment equations (\ref{MOMENTS}), while }$\widehat{%
\lambda }_{3}(1,t_{o})$ denotes\textit{\ }%
\begin{equation}
\widehat{\lambda }_{3}(1,t_{o})\equiv \widehat{\lambda }_{3}(\mathbf{r}_{1},%
\mathbf{v}_{1},t_{o})=\int\limits_{\Gamma _{1}}d\mathbf{x}_{2}\lambda _{3}(%
\mathbf{r},\mathbf{v,}t_{o})f_{1}(\mathbf{r}_{2},\mathbf{v}_{2},t_{o})
\label{solution for lambda_3}
\end{equation}%
and $\lambda _{3}(\mathbf{r},\mathbf{v,}t_{o})$ is determined in such a way
to fulfill the constraint (\ref{NEW CONSTRAINT-1}).

\textit{It follows that:}

\begin{enumerate}
\item $f_{1}(1,t_{o})$\textit{\ is strictly positive and determined as
velocity and phase-space moments the fluid velocity} $\mathbf{V}(\mathbf{r}%
_{1}\mathbf{,}t_{o}),$\textit{\ the kinetic pressure} $p_{1}(\mathbf{r}_{1}%
\mathbf{,}t_{o})$ \textit{as well as the thermodynamic entropy} $S_{T}$ \cite%
{Ellero2005,Tessarotto2006,Tessarotto2007b};

\item \textit{due to the arbitrariness of} $\left\langle
f_{1}(1,t_{o})\right\rangle _{\Omega },$ $f_{1}(1,t_{o})$ \textit{is
generally non-Gaussian and non-isotropic in velocity space;}

\item \textit{the PDF }$f_{1}(t_{o})$\textit{\ does not require the
specification of constraints on any higher-order multi-point PDF of order\ }$%
s>2.$
\end{enumerate}

PROOF

In fact there it results identically%
\begin{equation}
\delta \int\limits_{\Gamma _{1}}d\mathbf{x}h(t_{o})\frac{\left\langle
f_{1}(1,t_{o})\right\rangle _{\Omega }}{\left\langle h(1,t_{o})\right\rangle
_{\Omega }}\ln \frac{\left\langle f_{1}(1,t_{o})\right\rangle _{\Omega }}{%
\left\langle h(1,t_{o})\right\rangle _{\Omega }}=0.
\end{equation}%
From the Euler-Lagrange equation it is immediate to reach Eq.(\ref{solution
for h}) for $h(1,t_{o})$ [see the Euler-Lagrange (\ref{E-L equation})
above]. To prove 1) and 2) we notice that, by construction $f_{1}(1,t_{o})$%
\textit{\ }is strictly positive its moments satisfy the correspondence
principle constraints, while from (\ref{NON-Maxwellian}) and (\ref{solution
for h}) it follows that $f_{1}(1,t_{o})$ is generally non-Maxwellian.
Finally to prove 3) it is sufficient to notice that constraints of the type (%
\ref{NEW CONSTRAINT-1}) to be placed $s$-point PDFs (with $s>3$) leave
unchanged the Euler-Lagrange equation (\ref{E-L equation}). Q.E.D.

As a consequence of THM.1, Eqs. (\ref{NON-Maxwellian}) and (\ref{solution
for h}) provide the \textit{general solution} for the initial PDF $%
f_{1}(1,t_{o}),$ fulfilling all the physical realizability conditions
defined above [see Sec.3]. Remarkably, the solution determined in this way
\textit{does not require the specification of possible infinite set of
additional constraints}, analogous to Eq.(\ref{NEW CONSTRAINT-1}), \textit{%
to be placed on the initial multi-multipoint PDFs} \textit{of order} $s>2.$

\section{Particular solutions}

It is immediate to show that particular solutions are respectively provided
by:

a) the \textit{generally non-Gaussian PDF}:

\begin{equation}
f_{1}(1,t_{o})=\left\langle f_{1}(1,t_{o})\right\rangle _{\Omega }\frac{%
h(1,t_{o})}{\left\langle h(1,t_{o})\right\rangle _{\Omega }},
\label{caso part-1}
\end{equation}%
with $h(1,t_{o})$ the Gaussian PDF%
\begin{equation}
h(1,t_{o})=\exp \left\{ -1-\lambda _{o}(\mathbf{r}_{1o},t_{o})-\lambda _{2}(%
\mathbf{r}_{1o},t_{o})u(1,t_{o})^{2}\right\} ;  \label{caso part-1b}
\end{equation}

b)\textit{\ the Gaussian PDF}%
\begin{equation}
f_{1}(1,t_{o})=f_{M}(\mathbf{u}(\mathbf{r}_{1}\mathbf{,}t_{o});p_{1}(\mathbf{%
r}_{1}\mathbf{,}t_{o})),  \label{caso part-2}
\end{equation}%
with
\begin{equation}
f_{M}(\mathbf{u}(\mathbf{r,}t_{o});p_{1}(\mathbf{r,}t_{o}))=\frac{1}{\pi
^{3/2}v_{thp}^{3}(\mathbf{r}_{1}\mathbf{,}t_{o})}\exp \left\{ -\frac{\mathbf{%
u}(\mathbf{r}_{1}\mathbf{,}t_{o})^{2}}{v_{thp}^{2}(\mathbf{r}_{1}\mathbf{,}%
t_{o})}\right\} ,  \label{GAUSSIAN PDF}
\end{equation}%
and $v_{thp}(\mathbf{r}_{1}\mathbf{,}t_{o})=\sqrt{2p_{1}(\mathbf{\mathbf{r}}%
_{1},t_{o},\alpha )/\rho _{o}}$ denoting the thermal velocity associated to
the kinetic pressure.

\vskip5mm

\textbf{THM.2 - Particular solutions for} $f_{1}(t_{o})$

$f_{1}(t_{o})$ \textit{coincides either with (\ref{caso part-1}) or (\ref%
{caso part-2}) respectively if :}

A) \textit{there results identically}$:$%
\begin{equation}
\lambda _{3}(\mathbf{r}_{1}\mathbf{,v}_{1}\mathbf{,}t_{o})\equiv 0;
\label{constraint-0}
\end{equation}

B) \textit{besides (\ref{constraint-0}), also the equation\ }%
\begin{equation}
\frac{\left\langle f_{1}(t_{o})\right\rangle _{\Omega }}{\left\langle
h(t_{o})\right\rangle _{\Omega }}=1  \label{constraint}
\end{equation}

\textit{holds identically.}

PROOF

In validity of the constraint (\ref{constraint-0}) Eqs. (\ref{caso part-1})
and (\ref{caso part-1b}) follow from Eqs.(\ref{solution for h}) and (\ref%
{solution for lambda_3}). In validity of Eq.(\ref{constraint}) too, it is
immediate to prove that PEM implies\textit{\ }that for arbitrary variations $%
\delta h(\mathbf{r,u,}t_{o})$ it must result identically$:$
\begin{eqnarray}
&&\int\limits_{\Gamma _{1}}d\mathbf{x\delta }h(\mathbf{r,v,}t_{o})\left\{
1+\ln h(\mathbf{r,v,}t_{o})+\right. \\
&&\left. +\lambda _{o}\mathbf{+\lambda }_{2}\mathbf{u}^{2}\right\} =0.
\notag
\end{eqnarray}%
It follows that the Lagrange multipliers $\lambda _{o}(\mathbf{r,}t_{o})$%
\textbf{\ }and\textbf{\ }$\mathbf{\lambda }_{2}(\mathbf{r,}t_{o}),$
determined by imposing the correspondence principle, necessarily require
\begin{equation}
f_{1}(t_{o})=f_{M}(\mathbf{v-V}(\mathbf{r,}t_{o});p_{1}(\mathbf{r,}t_{o})).
\end{equation}

Q.E.D.

\section{Conclusions}

An axiomatic approach, based on the IKT-statistical model $\left\{
f_{1},\Gamma _{1}\right\} $, has been developed to determine the initial
condition for the 1-point PDF $f_{1}$ which characterizes an isothermal and
isentropic incompressible NS fluid.

In particular, we have proven that - extending the statistical approach
earlier developed \cite%
{Tessarotto2004,Ellero2005,Tessarotto2006,Tessarotto2007b,Tessarotto2008-4}
- the initial 1-point PDF can be uniquely determined by invoking the
principle of entropy maximization. The present theory is based on the
assumption that the initial PDF satisfies suitable physical realizability
conditions. \ In detail, this requires that the initial PDF $f_{1}(t_{o})$
be constructed in such a way to satisfy the physical requirements placed by :

\begin{itemize}
\item the correspondence principle [in particular Eqs.(\ref{MOMENTS})],
prescribing that the PDF determines in terms of suitable moments the
complete set of fluid fields;

\item the \ \textit{velocity-frequency density function }(V-FDF) $\widehat{f}%
_{1}^{(freq)}(\mathbf{v,}t_{o})$ \ (\ref{constraint on f1 (at initial time)}%
);

\item the $2$\textit{-point velocity difference-frequency density function}
(VD-FDF) $\widehat{f}_{2}^{freq}(\mathbf{r},\mathbf{v,}t_{o}),$ in term of
the constraint (\ref{NEW CONSTRAINT-1}) $.$
\end{itemize}

The theory has several important consequences:

\begin{enumerate}
\item THM.1 provides the general form of the initial 1-point PDF satisfying
the previous physical requirements;

\item the initial 1-point PDF is generally \textit{non-Gaussian} \textit{and
non-isotropic in velocity-space} (see THM. 1).

\item particular solutions include, including the Gaussian 1-point PDF, are
discussed in THM.2.
\end{enumerate}

In addition, the present theory provides an answer to the requirements
placed in Sec.1, namely that the initial PDF $f_{1}(t_{o})$ obtained in this
way is unique. In fact we have proven (THM.1) that the determination of $%
f_{1}(t_{o})$ does not require the specification of higher-order PDFs of
order\ $s>2$ and thus \textit{provides the general solution for the initial }%
$f_{1}(t_{o})$ \textit{of the} \textit{IKT-statistical model }$\left\{
f_{1},\Gamma _{1}\right\} .$

\section{Acknowledgments}
Work developed in cooperation with the CMFD Team, Consortium for
Magneto-fluid-dynamics (Trieste University, Trieste, Italy).
Research partially performed in the framework of the GDRE (Groupe
de Recherche Europeenne) GAMAS.



\end{document}